\documentclass[ twocolumn,tighten]{aastex631}
\usepackage{mathtools}
\usepackage{txfonts} 
\usepackage{hyperref}
\usepackage[utf8]{inputenc}

\newcommand{\ud}{\mathrm{d}}
\defcitealias{PaperI}{EHTC~I}
\defcitealias{PaperII}{EHTC~II}
\defcitealias{PaperIII}{EHTC~III}
\defcitealias{PaperIV}{EHTC~IV}
\defcitealias{PaperV}{EHTC~V}
\defcitealias{PaperVI}{EHTC~VI}
\defcitealias{PaperVII}{EHTC~VII}
\defcitealias{PaperVIII}{EHTC~VIII}
\newcommand{\m}{M\,87*}
\newcommand{\s}{Sgr\,A*}
\newcommand{\bhspin}{a_*}

\def\lsim{\mathrel{\raise.3ex\hbox{$<$\kern-.75em\lower1ex\hbox{$\sim$}}}}
\def\gsim{\mathrel{\raise.3ex\hbox{$>$\kern-.75em\lower1ex\hbox{$\sim$}}}}

\begin{document}                                              

\title{
Photon Ring Symmetries in Simulated Linear Polarization Images of Messier 87*
}
\shorttitle{Photon Ring Symmetries}

\author[0000-0002-7179-3816]{Daniel~C.~M.~Palumbo}
\affil{Center for Astrophysics $\vert$ Harvard \& Smithsonian, 60 Garden Street, Cambridge, MA 02138, USA}
\affiliation{Black Hole Initiative at Harvard University, 20 Garden Street, Cambridge, MA 02138, USA}

\author[0000-0001-6952-2147]{George~N.~Wong}
\affiliation{School of Natural Sciences, Institute for Advanced Study, 1 Einstein Drive, Princeton, NJ 08540, USA}
\affiliation{Princeton Gravity Initiative, Princeton University, Princeton, New Jersey 08544, USA}

\begin{abstract} 
The Event Horizon Telescope (EHT) recently released the first linearly polarized images of the accretion flow around the supermassive black hole Messier 87*, hereafter \m{}. The spiraling polarization pattern found in EHT images favored magnetically arrested disks (MADs) as the explanation for the EHT image. With next-generation improvements to very long baseline interferometry (VLBI) on the horizon, understanding similar polarized features in the highly lensed structure known as the ``photon ring,'' where photons make multiple half-orbits about the black hole before reaching the observer, will be critical to analysis of future images. Recent work has indicated that this image region may be depolarized relative to more direct emission. We expand this observation by decomposing photon half-orbits in the EHT library of simulated images of the \m{} accretion system and find that images of MAD simulations show a relative depolarization of the photon ring attributable to destructive interference of oppositely spiraling electric field vectors; this antisymmetry, which arises purely from strong gravitational lensing, can produce up to ${\sim}50\%$ depolarization in the photon ring region with respect to the direct image. In systems that are not magnetically arrested and with the exception of systems with high spin and ions and electrons of equal temperature, we find that highly lensed indirect sub-images are almost completely depolarized, causing a modest depolarization of the photon ring region in the complete image. We predict that next-generation EHT observations of \m{} polarization should jointly constrain the black hole spin and the underlying emission and magnetic field geometry.
\end{abstract}


\section{Introduction}\label{sec:intro}

In April 2019, the Event Horizon Telescope (EHT) released the first image of the accretion flow surrounding the Messier 87 supermassive black hole, hereafter \m{} \citep[][hereafter EHTC I-VI]{PaperI,PaperII,PaperIII,PaperIV,PaperV,PaperVI}. In April 2021, the EHT followed up with linearly polarized images (also at the 20 $\mu$as angular resolution of the EHT) and identified \m{}'s spiraling polarization structure as the imprint of poloidal magnetic fields threading the accretion flow \citep[][hereafter EHTC VII-VIII]{PaperVII,PaperVIII}. 
In this article, motivated by the steadily improving prospects for very long baseline interferometry (VLBI) using longer baselines and higher frequencies, we explore the polarimetric properties of the so-called ``photon ring'' in simulated images of \m{}. Here, we consider the photon ring to be the image region in which photons arrive at the observer after making one or more half-orbits about the black hole. 

The properties of polarized light emitted near a black hole have been well-studied for decades, especially in the context of x-ray polarization \citep{Connors_1977,Connors_1980}. More recently, efforts to describe the relationship between the local black hole accretion system parameters and the observed submillimeter image have been enabled by improvements in the analytic understanding of the problem. \citet{GL_null} and \citet{GL_lensing} form an excellent primer on the matter of analytic ray tracing in a Kerr spacetime, while \citet{Narayan_2021} and \citet{Gelles_Kerr} provide convenient toy models for understanding the physical origin of polarization signatures in the image domain.

In this article, we consider the polarization of direct and indirect lensed images of the accretion flow, for which \citet{Johnson_2020} and \citet{Himwich_2020} make the most relevant predictions. These papers predict symmetry relations between images formed from multiple orbits of photons around the black hole. This paper explores whether these symmetries are present in images produced from general relativistic magnetohydrodynamics (GRMHD) simulations.

 Recently, \citet{Alejandra_2021} observed that in images of GRMHD simulations of \m, the photon ring region  often has lower fractional polarization than the rest of the more weakly lensed ``direct'' image. We examine this result by decomposing the image into sub-images corresponding to successive photon half-orbits and by disentangling two types of depolarization: the destructive interference of oppositely polarized sub-images (which decreases total polarization and fractional polarization), and the summation of polarized and unpolarized sub-images (which decreases fractional polarization without affecting the total polarization). We focus our analysis on the initial library of GRMHD simulation snapshots produced for \citetalias{PaperV} and considered in \citet{PWP_2020}, hereafter P20.

While the intrinsic depolarization of a particular highly lensed sub-image is well-explained by \citet{Alejandra_2021}, we explain the destructive interference of polarized sub-images in the context of analytically motivated symmetries of the black hole spacetime using the rotationally symmetric Fourier mode ``$\beta_2$'' described in P20 and used extensively in \citetalias{PaperVIII}. This Fourier mode conveniently describes the appearance of spirals of polarization analogously to $E$ and $B$ modes in studies of the cosmic microwave background \citep[see, e.g.][]{Kamionkowski_2016}. The phase of the mode encodes whether the rotationally symmetric pattern of polarization vectors is directed radially, azimuthally, or in-between in a handed spiral.

As we will show, the appearance of the accretion flow is highly dependent on the underlying magnetization state, here either Standard And Normal Evolution (SANE) or a Magnetically Arrested Disk (MAD). In MAD flows, significant magnetic flux builds up on the black hole event horizon, increasing the magnetic pressure until it is large enough to counteract the inward ram pressure of the gas and intermittently halt the accretion; by contrast, SANE flows exhibit significantly weaker magnetic fields and thereby have steadier accretion \citep{Ichimaru1977, Igumenschchev_2003,Narayan_et_al_2003}. As discussed in P20 and \citetalias{PaperVIII}, MADs are associated with more radial and vertical magnetic fields (collectively, poloidal) in addition to having lower Faraday rotation depths than SANEs. 

In this letter, we will generally divide results along these magnetization states, black hole spin, and the $R_{\rm high}$ electron heating parameter introduced by \citet{Mosci_2016}, which sets the local ratio of electron to ion temperatures according to the ratio of gas and magnetic pressures. Generally $R_{\rm high}$ tunes the dominant emission region and Faraday rotation/optical depths; larger values correspond to greater off-midplane emission and higher rotation/optical depths. For a deeper discussion of the influence of $R_{\rm high}$, see the original text \citep{Mosci_2016} as well as  \citetalias{PaperV} and \citetalias{PaperVIII}.

In keeping with the current literature, we will label lensed accretion flow images by a number $n$ corresponding to the number of photon half-orbits between the observer and the termination point of the geodesic. The $n=0$ image thus corresponds to the weakly lensed direct image of the flow, which generally dominates all image metrics including the azimuthal symmetry modes discussed in P20. The $n=1$ image comprises most of the highly lensed emission frequently referred to as the ``photon ring'' in previous work. Any geodesics with support (that is, with a non-zero flux contribution) in the $n=1$ image will generally also contain flux in the $n=0$ image; the full image can be assembled by summing Stokes parameters over all sub-images. 

\begin{figure*}
    \centering
    \includegraphics[width=1.02\textwidth]{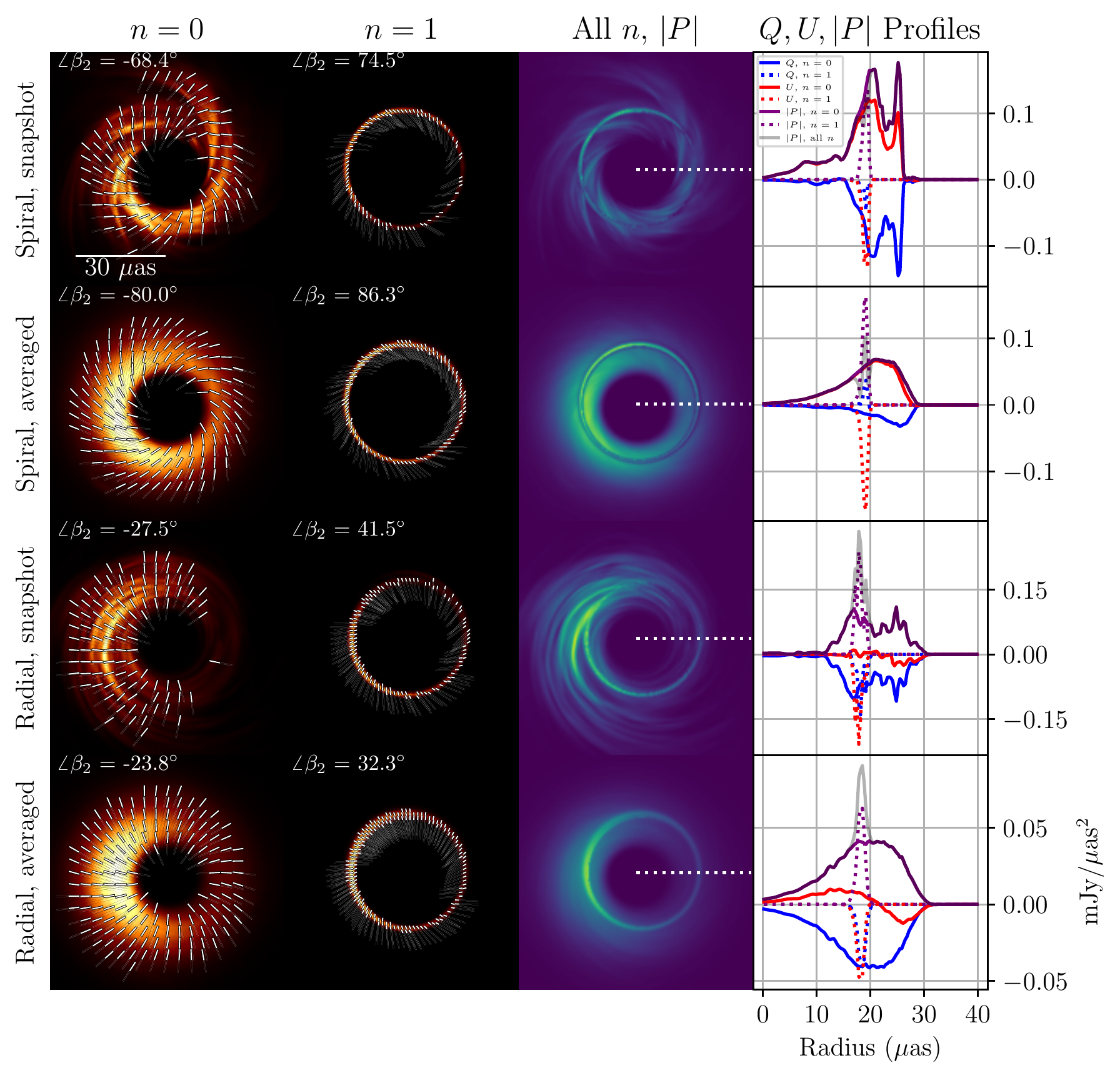}
    \caption{Top row, left column: random snapshot of the $n=0$ Stokes $I$ sub-image from a MAD simulation with $\bhspin=0.5$ and $R_{\rm high}=10$ with EVPA ticks shown where total intensity exceeds 1\% of its maximum and fractional polarization is at least 10\%. Second column: same as left column, but $n=1$. Third column: total linearly polarized flux $|P| = |Q+iU|$ of full image. Fourth column: profiles of Stokes $Q$, $U$, and $|P|$ along the profile depicted as a dotted white line in the third column. Second row: same as top row, but time averaged over 200 snapshots across quiescence. Bottom two rows: same as top rows, but for a SANE simulation with $\bhspin=0.94$ and $R_{\rm high}=1$. Simulations were chosen to have nearly imaginary and nearly real $\beta_2$ coefficients, respectively. EVPA spirals change handedness across sub-images. For accretion flows with more spiraling magnetic fields and thus less radially directed EVPA, this causes destructive interference near the photon ring, producing the dark depolarized edge in the MAD $|P|$, when direct and indirect images have comparable brightness. In the spiraling model, the sub-image partially depolarizes the total image, while in the more radial model, the sub-image 
    {strictly} increases total $|P|$ near the critical curve.}
    \label{fig:examples}
\end{figure*}

In \autoref{sec:pred} we provide simple predictions for symmetries of polarization across lensed sub-images in flat, Schwarzschild, and Kerr spacetimes with optically thin synchrotron emission. In \autoref{sec:GRMHD} we compute the polarization fraction and azimuthal symmetry metrics for the direct and highly lensed image components in the \citetalias{PaperV} image library. We conclude with a discussion in \autoref{sec:dis}.

\section{Analytic Predictions}
\label{sec:pred}

In order to describe spirals of polarization, we use azimuthal Fourier coefficients $\beta_m$ in the image domain as in P20:
\begin{align}
    \beta_m &=\dfrac{1}{I_{\rm tot}} \int\limits_{0}^{\infty} \int\limits_0^{2 \pi} P(\rho, \varphi) \, e^{- i m \varphi} \; \rho \mathop{\ud\varphi}  \mathop{\ud\rho},\\
    I_{\rm tot} &= \int\limits_{0}^{\infty} \int\limits_0^{2 \pi} I(\rho, \varphi) \; \rho \mathop{\ud\varphi} \mathop{\ud\rho}.
\end{align}
Here, $P=Q+iU$ is the complex polarization (with $Q$ and $U$ the usual Stokes parameters), $\rho$ is the image radial coordinate, $\varphi$ is the image azimuthal coordinate, and $I_{\rm tot}$ is the total Stokes $I$ flux in the image. The $\beta_2$ coefficient captures rotationally symmetric polarization structure.

It is instructive to consider the direct image of a particular accretion flow, which we shall embed in different spacetimes. For this purpose, we will use the spiral of the electric vector position angle (EVPA) shown for the time-averaged MAD in \autoref{fig:examples} (second row, first column). This pattern has $\beta_2 \sim -i$ and thus has phase $\angle\beta_2 \sim -\pi/2$. Recall that for synchrotron emission, which dominates \m{} emission at EHT wavelengths, the polarization is oriented perpendicular to the screen-projected magnetic field in the absence of lensing and Faraday effects.

If viewed from the opposite side, in flat space we would see a spiral of opposite handedness.
This flipped view is precisely what a photon sees upon half-orbiting the black hole once in the $n=1$ lensed sub-image in a Schwarzschild spacetime. If, however, the EVPA spiral were more akin to the bottom left of \autoref{fig:examples}, with $\beta_2\sim1$ (and thus $\angle\beta_2 \sim 0$), we would see that radial EVPA stays radial even when viewed from beneath. Thus when viewing a flow from nearly the top down, as we do for \m{}, the $\beta_2$ coefficient for the $n=1$ lensed sub-image should appear approximately complex conjugated.

This behavior can be derived analytically in a Schwarzschild black hole spacetime at face-on viewing inclinations through use of a simple toy model with a few additional simplifying assumptions. \citet{Narayan_2021}, hereafter N21, predicted the polarization structure produced by a thin, synchrotron-emitting ring of magnetized plasma orbiting a black hole. {N21 makes a number of simplifying assumptions about the magnetized fluid: first, that emission is confined to a ring centered on a Schwarzschild black hole, and second, that the fluid velocity and magnetic field are axisymmetric and offset by a constant angle, with equatorial components that are typically assumed to be exactly anti-aligned (though a vertical magnetic field component is permitted). The ray tracing is simplified as well; N21 traces rays to the equatorial plane analytically using an approximation provided by \citet{Beloborodov_2002}, which decreases in accuracy with increasing viewing inclination. Optical thinness is also assumed, removing the need for any radiative transfer in favor of a single geometric path-length factor, which does not affect polarization orientation.}

In appendix D of {N21}, 
they provide approximations for $\beta_2$ in terms of the axisymmetric fluid velocity and magnetic field orientation angles $\chi$ and $\eta$, emission radius $R$, and equatorial and vertical magnetic field components $B_{\rm eq}$ and $B_z$. To leading order in the fluid velocity $\beta$ and lensing factor $1/R$, they find:
\begin{align}
    \beta_{2,0} &= -\left(1-\frac{4}{R}\right)e^{2 i \eta}B_{\rm eq}^2 + \left(\frac{4}{R}e^{i \eta} - 2 \beta e^{i (\eta+\chi)}\right)B_{\rm eq} B_z,
\end{align}
{where we denote the $\beta_2$ coefficient for a particular sub-image $n$ by $\beta_{2,n}$.}

In order to compute $\beta_{2,n}$ in full generality, we would need to compute the radius of equatorial crossing for all $n$, which would lose the simplicity of the toy model by violating the assumptions of the underlying Beloborodov approximation made by 
{N21} \citep{Beloborodov_2002}. We instead make further approximations; for photons near the ``critical curve'' (that is, in Schwarzschild, those photons with screen radius $\rho$ near $\rho_c=\sqrt{27}${\,M}), the emission radius is essentially constant over $n$ as photons wind nearly circularly about the black hole approaching $R=3${\,M}. These circular orbits all impact the equatorial emitting ring at right angles; thus, for all sub-images, the emission angle with respect to the Boyer-Lindquist radius is simply $\alpha_G=\pi/2$. This assumption is worst for the $n=0$ image and essentially eliminates any contribution from vertical fields. {Thus, our predictions allow for any axisymmetric magnetic field configuration, although the prediction accuracy decreases as the field becomes more vertical with respect to the midplane.}

Finally, we make three additional substitutions to compute the azimuthal symmetry of the $n^{\rm th}$ sub-image. For the simple, circular photon orbits of the Schwarzschild spacetime, the Boyer--Lindquist azimuthal coordinate $\phi$ swaps sides on each subsequent half-orbit. Furthermore, the winding angle $\psi$ must increment by $\pi$ with each successive half-orbit. Finally, the photon angular momentum $k$ with respect to Boyer--Lindquist polar coordinate $\theta$ must swap sign each time the photon punctures the midplane in the N21 formalism. Thus, we take
\begin{align}
    \alpha_{G,n} &= \frac{\pi}{2},\\
    \phi_n &= \phi_0+n \pi,\\
    \psi_n &= \psi_0+n \pi,\\
    k_{\theta,n} &= (-1)^n k_{\theta,0}. 
\end{align}

Finally, we compute the full polarization prediction as a function of image azimuthal angle $\varphi$ and enforce a purely rotationally symmetric polarization by taking a face-on viewer (that is, the viewing inclination $\theta_o\rightarrow 0$). We take $\eta=\chi+\pi$ in keeping with the ``plasma drags field'' assumption taken in {N21}. 
We compute the complex polarization $P$ as a function of $n$ and find
\begin{align}
    P_n(\varphi) &= \beta_{2,n} e^{i 2 \varphi},\\
    \beta_{2,n} &= - e^{i(-1)^n 2 \eta}\\
    &= |\beta_{2,0}| e^{i (-1)^n \angle{\beta_{2,0}}},\\
    \rightarrow |P_n(\varphi) + P_{n+1}(\varphi)| &\propto {\, \rm real}(\beta_{2,0}).
    \label{eq:sum}
\end{align}
In simple terms: for face-on viewing of a flow around a Schwarzschild black hole, the $\beta_2$ coefficient 
{asymptotically approaches complex conjugation across sub-images for large $n$} and {approximately obeys this relation for small $n$ }
for emission near the critical curve, producing a depolarization that depends on the phase of $\beta_2$ in image regions with overlapping sub-images of adjacent index $n$. This prediction may be viewed as a zeroth order approximation for sub-image polarization in the {N21} model.

In a Kerr black hole spacetime, this conjugation structure does not apply so simply to $\beta_2$. This is unsurprising, since a spinning black hole imposes a twisting structure on spacetime that is flipped in handedness across the Boyer--Lindquist midplane. This problem has already been studied in detail by \citet{Himwich_2020}, hereafter H20. In both H20 and 
{N21}, conservation of the Walker--Penrose constant $\kappa$ is used for parallel transport of rays from the observer to the black hole \citep{Walker_Penrose_1970}. 
$\kappa$ is a property of individual geodesics that encodes the orientation of the polarization vector; for axisymmetric emission from the midplane, $\kappa$ itself approaches rotational symmetry across geodesics in the observer screen as the observer approaches face-on viewing (as is indicated later by \autoref{fig:madmatch}). {Even along a single geodesic, subsequent midplane crossings yield distinct values of $\kappa$ since the angle between the magnetic field and the geodesic differs between crossings. In essence, $\kappa$ is a conserved quantity only for a geodesic that emerges from a single, infinitesimal emitter and otherwise propagates through a vacuum; sub-images with distinct $n$ will yield distinct values of $\kappa$ because the emitting surface is itself distinct, even if the position of arrival on the observer screen is unchanged. This behavior differs from the closely related case of the Carter integral, which is constant along a geodesic regardless of the differing angles between the emission direction and magnetic field at the geodesic origin at each $n$ \citep{Carter1968}.} 

H20 found that near the critical curve, adjacent sub-images exhibit complex conjugation in $\kappa$. This 
{behavior is approached at large $n$}; for a single geodesic that encounters the midplane twice (that is, with support in $n=0$ and $n=1$), we expect approximate complex conjugation in the value of $\kappa$ corresponding to the $n=0$ midplane intersection and the $n=1$ midplane intersection.

The emission location, observer screen position, and magnetic field geometry specify $\kappa_n$, which in turn specifies polarization. The relationship between these parameters and the observed emission is the province of N21 and the extension to Kerr provided by \citet{Gelles_Kerr}, hereafter G21; we are more interested in the observable predicted by H20 regarding sums of successive sub-images:
\begin{align}
    |P_n(\varphi) + P_{n+1}(\varphi)| &\propto {\, \rm real}(\kappa_n)\approx{\, \rm real}(\kappa_{n+1}).
    \label{eq:kappa}
\end{align}
Evidently in Kerr, the conjugation symmetry of $\beta_2$ is promoted to one of $\kappa$, as the axisymmetries of the black hole connect geodesic properties to rotationally symmetric features in the observer screen. In each case, we expect maximum polarization in full (summed over $n$) images when the direct image value of $\beta_2$ or $\kappa$ is mostly real, and we expect maximum depolarization when the value is mostly imaginary. Note that taking dimensionless black hole spin $\bhspin \rightarrow 0$ and $\theta \rightarrow 0$ guarantees $\kappa_{n+1} = \kappa_n^*$ for all rays with support in sub-images $n$ and $n+1$, but for more general configurations this statement is {approximate, with increasing accuracy at larger $n$.}

\begin{figure*}
    \centering
    \includegraphics[width=0.99\textwidth]{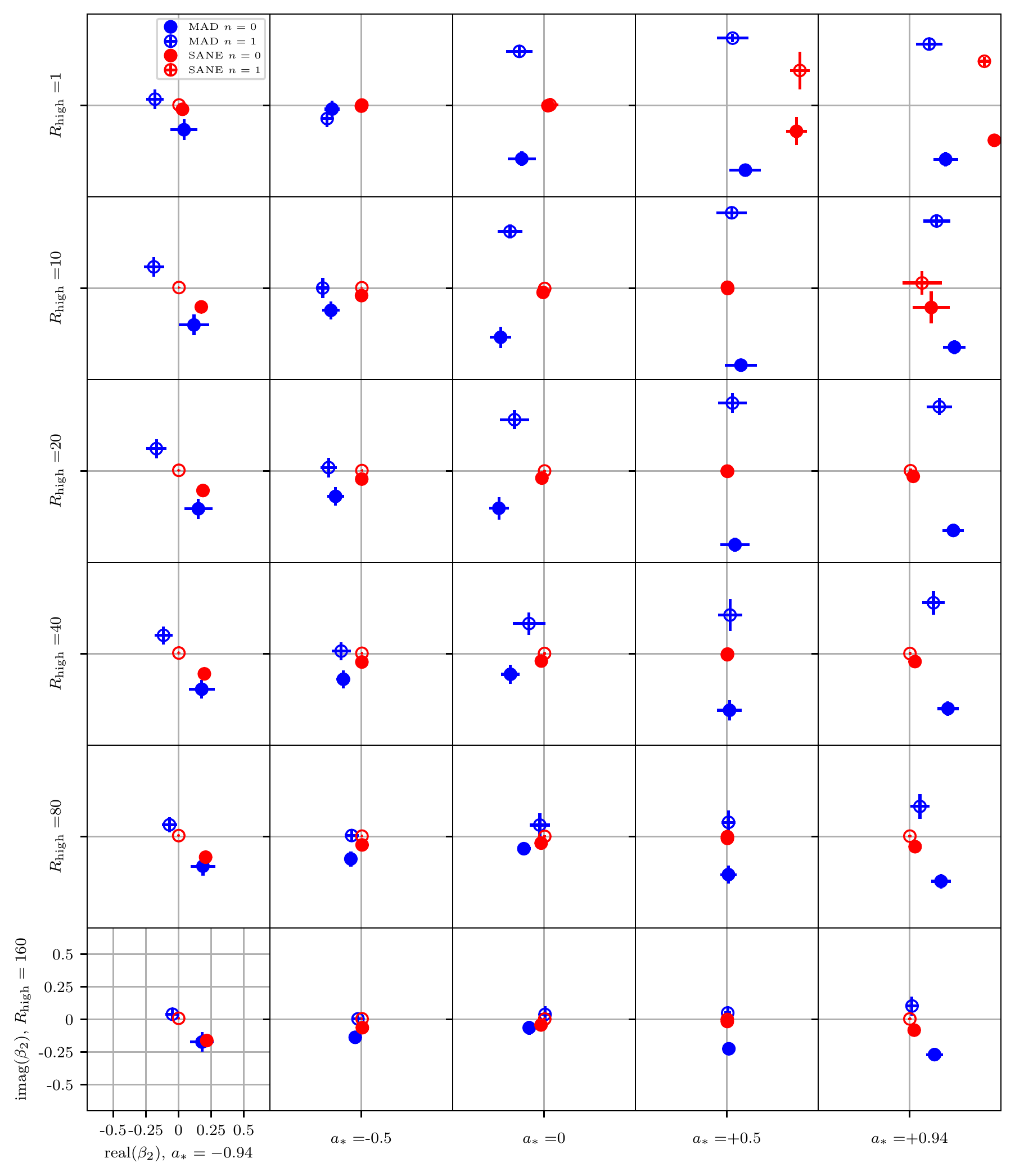}
    \caption{Values of the complex rotational symmetry coefficient $\beta_2$ for $n=0$ and $n=1$ sub-images across MAD and SANE simulations of varying spin and $R_{\rm high}$. Scale is constant across the grid; non-axial grid lines are hidden for clarity.}
    \label{fig:beta2s}
\end{figure*}

Formally, both \autoref{eq:sum} and \autoref{eq:kappa} are good approximations only for high $n$ in completely optically thin flows. We now summarize the predictions and assumptions for images of simulated flows, which are dominated by low-$n$ emission:
\begin{enumerate}
    \item We expect complex conjugation between $\beta_{2,0}$ and $\beta_{2,1}$ if emission is viewed nearly face-on, the observer screen radius of the $n=0$ image is near the critical curve, vertical fields in the emission region are weak, and the black hole has low spin.
    \item We expect the same restrictions for complex conjugation between $\kappa_0$ and $\kappa_1$, but allowing for arbitrary spin.
    \item We expect deviations from this behavior in realistic simulations because images at the horizon scale are dominated by low $n$ structures which {only weakly }
    obey symmetries relying on perpendicular impact with an equatorial emission surface.
    \item We expect large deviations from both \autoref{eq:sum} and \autoref{eq:kappa} in models for which there is prominent emission at large radii or far from the equatorial plane.
    \item We expect both \autoref{eq:sum} (if Schwarzschild) and \autoref{eq:kappa} (if Kerr) to hold for arbitrary emission and magnetic field geometries for large $n$, assuming zero optical depth.
\end{enumerate}

In the next section, we will analyze the GRMHD library largely through the lens of $\beta_2$, as $\beta_2$ is immediately observationally accessible (given some notion of the image center), while $\kappa$ is a geodesic property requiring knowledge of the underlying astrophysics. In \autoref{sec:dis}, we will touch on whether nearly imaginary $\kappa$ can produce destructively interfering sub-images with non-conjugating $\beta_2$ in order to understand trends in the GRMHD simulations.

\section{GRMHD Results}
\label{sec:GRMHD}

We now explore whether the sub-image symmetries are present in simulated images of the \m{} accretion flow. We use the GRMHD simulations used for the Stokes $I$ analysis of \citetalias{PaperV}, which were simulated with {\tt{}iharm3D} \citep{Gammie_HARM_2003, IHARM3d_prather} and ray traced with {\tt{}ipole} \citep{IPOLE_2018}. This library consists of five dimensionless black hole spins (-0.94, -0.5, 0, 0.5, 0.94), two magnetization states (MAD and SANE), and six values of the electron heating parameter $R_{\rm high}$ (1, 10, 20, 40, 80, 160). 
{The images produced for this paper are for a black hole with mass $= 6.5 \times 10^{9}\,M_{\odot}$ at a distance of $16.8\,$Mpc. These numbers were chosen to match the result of the EHT analysis, although they produce a slightly increased angular size compared to the fiducial image set produced in \citetalias{PaperV}. This change does not noticeably affect any quantities we compute. More details about the simulation pipeline can be found in \citet{Wong_2022}.}
The viewing inclination is chosen so that the black hole spin vector is always oriented away from the viewer at an angle of $17^\circ$ from the screen normal, irrespective of the motion of the accretion disk at large scales. This choice preserves the direction of rotation of the fluid in the inner accretion flow across all models (see Figure 5 in \citetalias{PaperV} and \citet{Wong_2021}). We have performed the ray tracing anew, using 200 images across the periods of quiescence identified in \citetalias{PaperV} and decomposing emission into $n=0$ and $n=1$ sub-images at a resolution of $1/3$\, $\mu$as. This decomposition follows the scheme in \citet{Johnson_2020}, in which sub-images are delineated by turning points in Boyer--Lindquist $\theta$ along photon geodesics.

In \autoref{fig:examples}, we examine a MAD with $\bhspin=0.5$ and $R_{\rm high}=10$ and a SANE with $\bhspin=0.94$ and $R_{\rm high}=1$. We choose these two models because they lead to spiraling and radial EVPA patterns, respectively. The profiles in the right column of the figure show patterns in polarized flux with a few notable features consistent with our analytic intuition: first, the spiraling model appears depolarized in some regions near the photon ring, while the radial model has a strictly increased polarized flux profile. In the spiraling model, where the peak brightness of the photon ring exceeds the direct image brightness, we see a pair of dark bands surrounding a brighter ring of polarization. These dark bands correspond to the points in the $n=1$ profile where $P_{n=1}$ is equal and opposite to $P_{n=0}$; where $P_{n=1}$ is at its maximum, the total polarization can exceed the $n=0$ profile even if the overall structure exhibits depolarization, as seen in the second row, fourth column of \autoref{fig:examples}. Ultimately, the flux ratio between the $n=0$ and $n=1$ image determines whether a single dark photon ring region or two dim bands are present in $|P|$. Note that two types of depolarization can be present in general: one in which an unpolarized and polarized image are summed, decreasing fractional polarization $m$ but not total polarized flux $|P|$, and another in which polarization interferes destructively, decreasing both $m$ and $|P|$. The dark bands in $|P|$ are clear evidence of the latter type.

We now turn to analyzing the full set of $n=0$ and $n=1$ images for the GRMHD image library introduced earlier. We compute the $\beta_2$ coefficient with one major difference from P20 -- we do not blur the image by the nominal 20\,$\mu$as resolution of the 2017 \m{} EHT data. We are not interested in the bulk properties of the flow that can be inferred from a blurry image. Rather, we aim to identify whether or not sub-images exhibit near-conjugation symmetries in the $\beta_2$ coefficient, and the blurring procedure is orthogonal to this consideration. In any case, blurring generally depolarizes the image and suppresses high-frequency modes in the azimuthal Fourier decomposition; the effect of a 20 $\mu$as blur on $\beta_2$ is to slightly reduce amplitude while leaving phase essentially unchanged.

\autoref{fig:beta2s} shows the complex $\beta_2$ coefficients for each set of simulation and image parameters for both $n=0$ and $n=1$. Recalling from \citetalias{PaperVIII} that the $\beta_2$ amplitude is essentially redundant with the average linear polarization fraction in an image, we note that any set of images with nearly zero $|\beta_2|$ is also nearly completely depolarized. Thus, we observe several general trends:
\begin{enumerate}
    \item Increasing $R_{\rm high}$ weakly depolarizes the $n=0$ image and greatly depolarizes the $n=1$ image, particularly in MADs.
    \item Images with significant fractional polarization in the $n=1$ component exhibit nearly conjugated $\beta_2$ coefficients between $n=0$ and $n=1$, with the largest deviation from this behavior occurring for retrograde MADs, particularly those with $\bhspin=-0.94$.
    \item In all cases except the $R_{\rm high}=1$ prograde spinning SANEs, SANE $n=1$ images are almost completely depolarized.
\end{enumerate}
These results follow trends identified in \citetalias{PaperVIII}: higher $R_{\rm high}$ depolarizes images overall and SANEs are typically more innately magnetically scrambled and are less polarized. Moreover, higher values of $R_{\rm high}$ correspond to stronger emission from the jet as opposed to the accretion flow. Emission from further out from the equatorial plane disobeys the assumptions required for the H21 symmetries, which may also contribute to violations of the equatorial emission predictions (though Faraday effects likely dominate).

\begin{figure*}[htbp]
    \centering
    \includegraphics[width=0.99\textwidth]{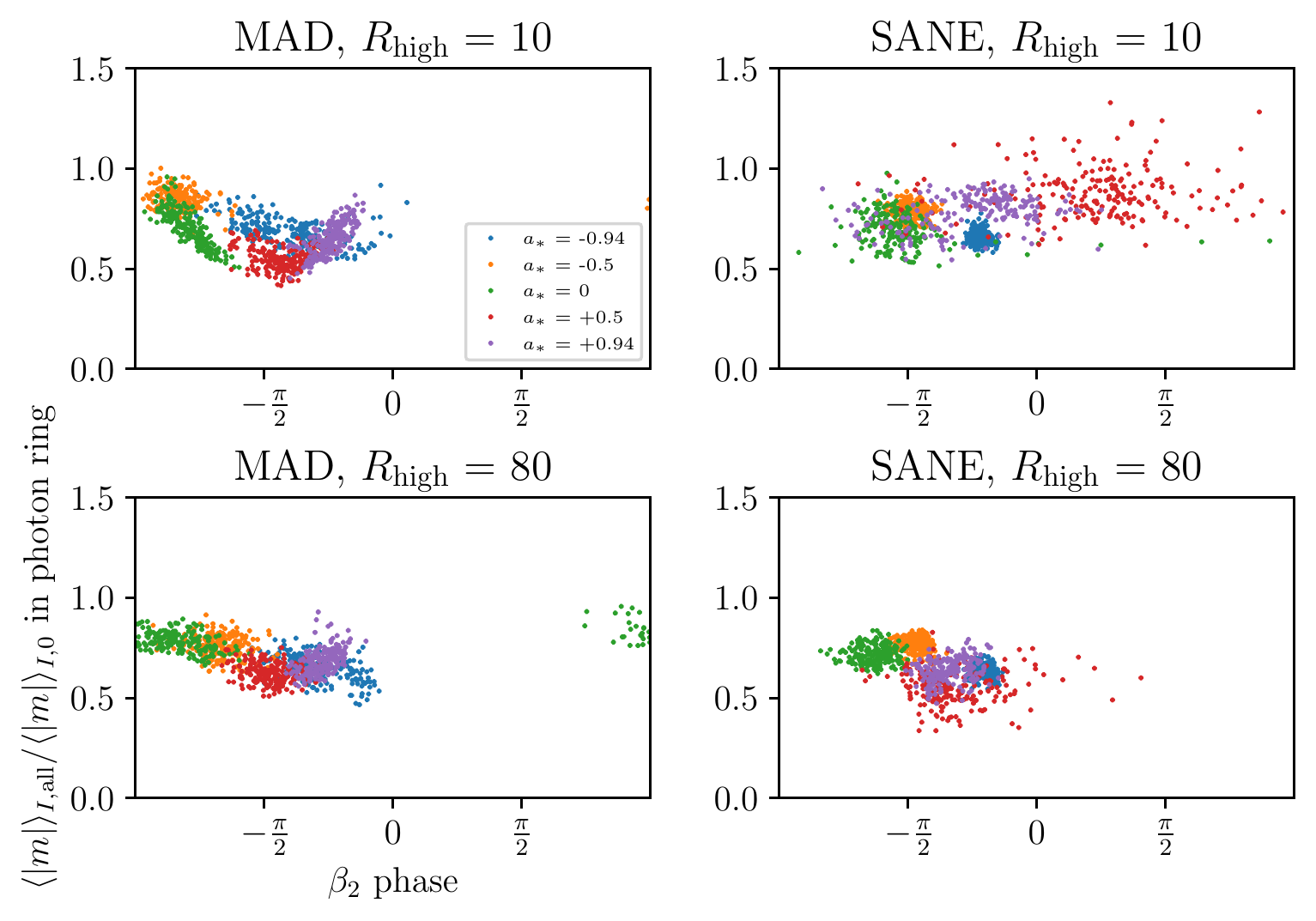}
    \caption{Top left: ratio of average fractional polarization in the all-$n$ image to that in the $n=0$ image  plotted against phase of the rotationally symmetric $\beta_2$ coefficient for images from a MAD model with $R_{\rm high} = 10$. Right: same as left, but for SANE images. Bottom row: same as top rows, but for $R_{\rm high} = 80$. 
    Images of prograde MADs exhibit sinusoidal behavior with respect to $\angle\beta_2$ due to destructive interference of oppositely handed EVPA spirals, while SANEs appear depressed by a roughly constant factor.}
    \label{fig:depol}
\end{figure*}

The retrograde high spin MAD {results} are peculiar in that 
{$\beta_{2,0}$ and $\beta_{2,1}$} are almost exactly opposite, reflecting an overall sign flip rather than a complex conjugation. Recall, however, that for general viewing angles, spins, magnetic fields and emission geometries, we only expect {approximate} complex conjugation of the underlying $\kappa$, 
not $\beta_2$. Although $\kappa$ can not be extracted simply from a polarized image, we will discuss later an example fluid configuration that can give rise to the MAD $\bhspin=-0.94$ pattern.

Next, we examine depolarization near the critical curve. Rather than set a mask based on possible photon ring locations as in \citet{Alejandra_2021}, we simply generate a mask using our ray traced $n=1$ images. Within the $n=1$ image, wherever the Stokes $I$ flux exceeds 1\% of its peak value, we consider the pixel to be in the ``photon ring region.'' We use this mask on the full, all-$n$ image to select photon ring pixels and compute the Stokes $I$-weighted fractional polarization as in \citetalias{PaperVII}. The details of these masking choices are not particularly relevant to the signal we are seeking: qualitatively, the full image should depolarize according to \autoref{eq:sum} for face-on, Faraday/optically thin emission. As we have seen in \autoref{fig:beta2s}, these assumptions are violated for almost all SANEs, as the indirect emission is almost completely depolarized. Thus, our expectation is that the average Stokes $I$-weighted fractional polarization near the critical curve in MADs should obey $\langle|m_{\rm avg}|\rangle_I \propto 1-\sin|\angle\beta_2|$, since more spiraling EVPA patterns cancel with their sub-image. In SANEs, where the $n=1$ image is generally depolarized, the polarized $n=0$ image is added to an unpolarized $n=1$ image, causing a net decrease in average fractional polarization. This decrease does not depend on $\angle\beta_2$, because the $n=1$ image is not polarized.

\begin{figure*}[htbp]
    \centering
    \includegraphics[width=0.99\textwidth]{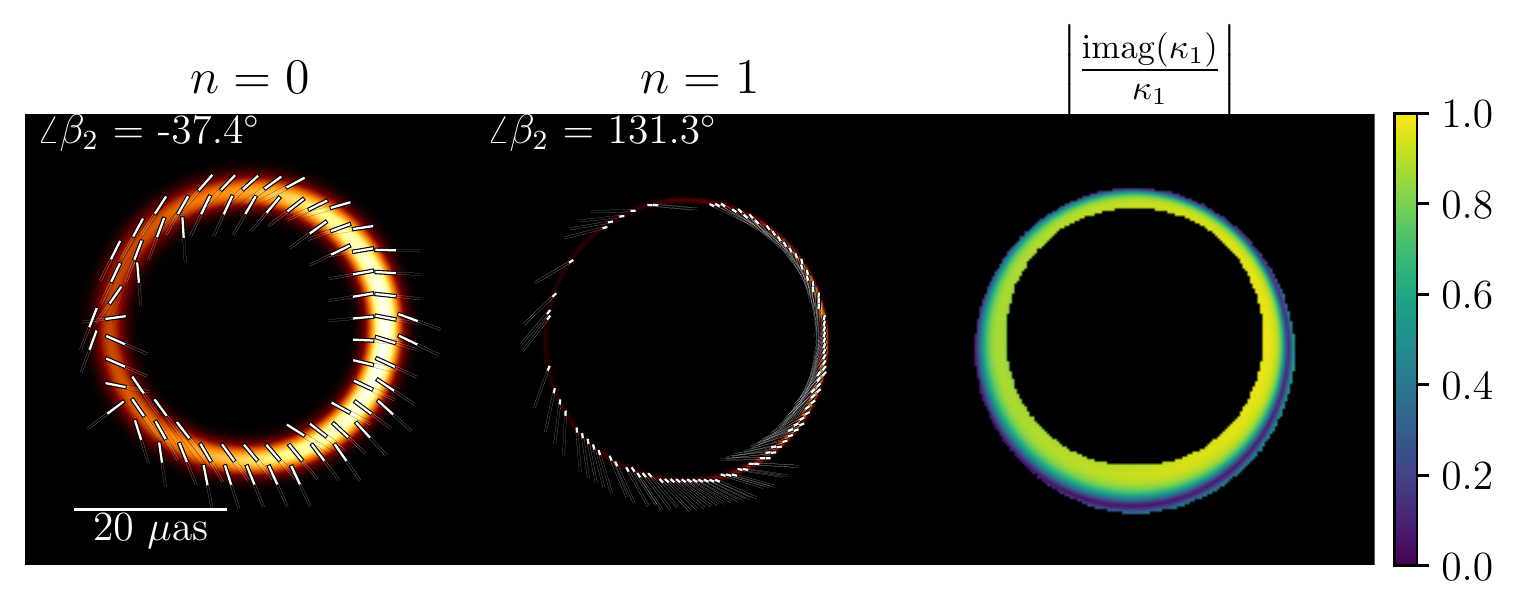}
    \caption{Example \citet{Gelles_Kerr} toy model for a synchrotron-emitting ring of polarized gas chosen to produce the $\beta_2$ phase offset seen in MADs with $\bhspin=-0.94$ in \autoref{fig:beta2s}. Left two columns: predicted $n=0$ and $n=1$ images shown with same plotting parameters as \autoref{fig:examples}. Right column: fractional imaginary part of the Penrose--Walker constant $\kappa$ underlying the predicted images (brighter corresponds to more imaginary). The nearly $180^\circ$ rotation in $\angle \beta_2$ corresponds to a nearly $90^\circ$ uniform rotation in EVPA between $n=0$ and $n=1$ as predicted by \citet{Himwich_2020} for (nearly) purely imaginary $\kappa$.}
    \label{fig:madmatch}
\end{figure*}

\autoref{fig:depol} shows the full-image-to-sub-image ratio of the Stokes $I$ weighted fractional polarization $\langle|m|\rangle_I$ for MADs and SANEs at $R_{\rm high} = 10$ and $80$ (corresponding to the second and fifth rows of \autoref{fig:beta2s}). If the mechanism for depolarization is indeed the destructive interference between oppositely handed spirals, the depolarization between the $n=0$ and all-$n$ images should be largest at $\angle\beta_2 = -\pi/2$; this is very visible as a ``U'' shape in the top left panel of the figure. At high $R_{\rm high}$ values, however, the sub-image is significantly depolarized, and adding a depolarized background to a polarized foreground weakens the dependence on $\angle\beta_2$, as can be seen in the bottom left. When the sub-image is most polarized and oppositely spiraling, as in the $R_{\rm high}=10$ MADs with $\bhspin = 0.5$, this destructive interference reduces the fractional polarization in the photon ring region by 50\% (a result that is sensitive to our choice of photon ring mask criterion). Meanwhile, the SANEs fall roughly in a horizontal line regardless of $R_{\rm high}$, indicating that the spiral interference is a negligible contribution to the depolarization.

As an aside, most images --- including nearly all MADs --- are confined to negative values of the $\beta_2$ phase, corresponding to a strict adherence to the assumption made in N21, i.e., that the magnetic field trails the fluid velocity. For a clockwise flow on the sky (here, dictated by our choice of viewing inclination), this assumption requires $-\pi<\angle\beta_2<0$. Prograde $\bhspin=0.5$ SANEs with $R_{\rm high} = 10$, however, are so close to the origin in the complex $\beta_2$ plane that the phase of $\beta_2$ wanders significantly even at quiescence. 

We conclude then that the dominant depolarization mechanism in the photon ring regions of MADs (especially at low $R_{\rm high}$) is the cancellation of EVPA spirals across sub-images, whereas for SANEs and MADs at very high $R_{\rm high}$ the mechanism is more likely the addition of an unpolarized $n=1$ image. Note that this effect is distinct from the depolarization of any individual sub-image, for which optical depth and Faraday effects governed by magnetic fields and electron temperature and density will be dominant; see \citet{Alejandra_2021} for a discussion of intrinsic depolarization of the long geodesics in the photon ring.

\section{Discussion}
\label{sec:dis}

We have computed the rotational symmetry coefficient $\beta_2$ for $n=0$ and $n=1$ sub-images from a large library of GRMHD simulations and found clear evidence for complex conjugation of the coefficient in almost all MADs and almost no SANEs. We found that the relative depolarization of the photon ring with respect to other image regions is greatest in MADs with $|\angle\beta_2|\sim\pi/2$. The almost universal depolarization of the $n=1$ image in SANE simulations suggests that any depolarization of the photon ring region comes from the sum of polarized and unpolarized sub-images rather than destructive interference. Collectively, we reproduce the result in \citet{Alejandra_2021} that the photon ring is often depolarized relative to the rest of the image and find a compelling case that, at low values of $R_{\rm high}$, this relative depolarization is primarily due to lensing symmetries. We attribute deviation from the complex conjugation of $\beta_2$ to prominent effects from spin, optical and Faraday rotation depth, and emission originating at large radii or far from the equatorial plane of the black hole.

We have found that retrograde highly spinning MADs (of any $R_{\rm high}$) do not exhibit obvious complex conjugation of $\beta_2$ (see the left column of \autoref{fig:beta2s}). As is apparent in \autoref{fig:beta2s} and \autoref{fig:depol}, these models are still heavily depolarized by the destructive interference of the $n=1$ image, but here 
{$\beta_{2,1}$} is nearly the negation of the direct image value. We investigate this behavior by reproducing it using the G21 toy model evaluated over an observer screen; we  use the Kerr Bayesian Accretion Modeling (\texttt{KerrBAM}) code \citep{KerrBAM} to generate images. We find that a broad class of models can reproduce the MAD $\bhspin=-0.94$ behavior when supplied with reasonable model parameters for such a flow. Non-zero inclinations, strong vertical fields, and emission at radii far from $R\approx \sqrt{27}-1$ {\,M} can all create large deviations from the conjugation symmetry. As one example, \autoref{fig:madmatch} considers a model of an emissive ring viewed at $\theta_o=17^\circ$ with a Gaussian profile centered at 3 M with a full width at half-maximum of 1 M and a predominantly radially infalling gas moving at $\beta=0.6$ with a non-negligible vertical field ($\chi=-7\pi/8$, $\eta=\chi+\pi$, $B_z=0.5$ in the terminology of {N21}. 
Here, distances are expressed in scale-free units, where $1 {\rm \,M} = \,{G M}/{c^2}$ for the gravitational constant $G$, black hole mass $M$, and speed of light $c$. We find a nearly $90^\circ$ offset in EVPA ($\sim180^\circ$ in $\angle\beta_2$) between the $n=0$ and $n=1$ sub-images, corresponding to the half-rotation in the complex plane seen in the $\bhspin=-0.94$ MADs. As expected, the underlying $\kappa$ is almost purely imaginary, with the exception of a narrow band far from the critical curve, giving rise to destructive interference through \autoref{eq:kappa}.

This replication is notable because the toy model is purely equatorial; we expect that in addition to the emission radius and magnetic field caveats, three-dimensional emission geometries (as seen in Figure 4 of \citetalias{PaperV}) would also cause a breakdown of the simple sub-image relations. \citet{Chael_2021} found that in time-averaged simulations of MAD accretion flows, emission is concentrated in a relatively thin region about the equatorial plane, which may mitigate violations of these simple symmetries. However, as we see in \autoref{fig:madmatch}, complicated three-dimensional geometry is not required to produce relations between sub-image polarization patterns that violate simple flat space intuition. It is thus remarkable that so many of the GRMHD models obey the simpler \autoref{eq:sum} relation despite the complicated structure of the spacetime geometry and emission features; we attribute this to the conclusion in \citetalias{PaperV} that the majority of emission in the GRMHD library tends to come from radii near 4.5\,M. As is evident from the ``just add one'' approximation that the lensed screen radius of emission from a radius $R$ appears nearly at $\rho \approx R+1$ for face-on viewing geometries, emission from near $R=4.5$\,M will fall near the critical curve, which is the most impactful requirement for applying \autoref{eq:sum} and \autoref{eq:kappa}. Moreover, we expected strong vertical fields to create large discrepancies from \autoref{eq:sum}; however, as touched on in \citet{PaperVIII} and \citet{Narayan_2021}, vertical magnetic fields near the equatorial plane in GRMHD (and likely in \m{}) tend to be fairly weak.

In this letter, we have only considered linear polarization. Recently, \citet{Moscibrodzka_2021} and \citet{Ricarte_2021} found that the photon ring region in images of GRMHD simulations often has the opposite sign of circular polarization compared to the direct image of the flow.  Although this signature appears similar to the lensing symmetries discussed here, the circular polarization antisymmetry is due to subtle details of the underlying magnetic field and its twist, and is a less generic property of roughly disk-like flows about a black hole. Moreover, since the 
{N21}/G21 toy model we consider in this paper does not consider polarized radiative transfer, it is insufficient for a detailed study of circular polarization, so more work remains in generating as intuitive a picture for circular polarization as has been done for linear polarization.

Given the conclusion of \citetalias{PaperVIII} that \m{} is likely magnetically arrested, we expect that future high-resolution VLBI measurements of the \m{} photon ring should find slight depolarization near the critical curve due to \m's gently spiraling EVPA pattern and low overall polarization. Further, the so-far consistent story that \m{}'s compact 230\,GHz emission is from somewhere in the 3 M to 5 M range of radii with weak vertical fields favors the application of the symmetries identified here and in \citet{Himwich_2020}. This is of particular relevance to the ngEHT (see, for example, \citet{Raymond_2021}), for which a primary objective is polarized studies of the \m{} $n=1$ sub-image. The variety of $\beta_2$ phase offset structures in \autoref{fig:beta2s} suggests that a measurement of the $n=0$ and $n=1$ polarization spirals would simultaneously constrain multiple properties of the black hole accretion system. While the black hole spin is constrained through the shape of the photon ring itself, the astrophysical details are inextricably linked to the spacetime through $\kappa$ and encoded in the relation between $n=0$ and $n=1$ polarization, meaning that model-fitting the magnetic field and plasma around the black hole might pin down the fluid velocity, magnetic field geometry, and the location of the emission in addition to parameters of the spacetime. We remain hopeful that within the decade, we will have a formal, if imprecise, constraint on the three-dimensional emission and magnetic field geometry of \m{}.

\begin{acknowledgements}
We thank Elizabeth Himwich, Ramesh Narayan, Monika Mo{\'s}cibrodzka, Zachary Gelles, Angelo Ricarte, and Michael Johnson for many helpful early discussions, as well as Thomas Bronzwaer for a thorough review of early drafts. {We also thank our reviewer for their thoughtful comments on our manuscript.} This work was supported by the Black Hole Initiative at Harvard University, which is funded by grants from the John Templeton Foundation and the Gordon and Betty
Moore Foundation to Harvard University. D.C.M.P. was supported by National Science Foundation grants AST 19-35980 and AST 20-34306. G.N.W.~gratefully acknowledges support from the Institute for Advanced Study.
\end{acknowledgements}

\software{\texttt{eht-imaging} \citep{Chael_closure},
          \texttt{ipole} \citep{IPOLE_2018},
          \texttt{KerrBAM} \citep{KerrBAM},
          Matplotlib \citep{matplotlib},
          Numpy \citep{numpy}}
\bibliography{references}

\end{document}